\documentclass[aps,prx,reprint,floatfix,amsmath,amssymb,superscriptaddress,footinbib]{revtex4-1}

\usepackage{graphicx}
\usepackage{epstopdf}
\usepackage{color}
\usepackage{hyperref}
\usepackage{bm}
\usepackage{printlen}
\usepackage{units}
\usepackage{esint}

\renewcommand{\vec}[1]{\mathbf{#1}}
\newcommand{\Torder}{\text{T}_{\tau} }
\newcommand{\Trace}{\text{Tr}\,}

\renewcommand{\Re}{\text{Re}}
\renewcommand{\Im}{\text{Im}}
\newcommand{\eps}{{\varepsilon}}

\hypersetup{hidelinks}

\newcommand{\Ham}{\mathcal{H}}

\newcommand{\VV}{\mathcal{V}}

\begin{document}

\title{Spectral sum rules reflect topological and quantum-geometric invariants}

\author{Alexander Kruchkov}

\affiliation{Department of Physics, Princeton University, Princeton, New Jersey 08544, USA}

\affiliation{Institute of Physics, {\'E}cole Polytechnique F{\'e}d{\'e}rale de Lausanne,  Lausanne, CH 1015, Switzerland, and Branco Weiss Society in Science, ETH Zurich, Zurich, CH 8092, Switzerland}

\affiliation{Department of Physics, Harvard University, Cambridge, Massachusetts 02138, USA}

\author{Shinsei Ryu}

\affiliation{Department of Physics, Princeton University, Princeton, New Jersey 08544, USA}

\date \today

\begin{abstract}
Topological invariants are fundamental characteristics reflecting global properties of quantum systems, yet their exploration has predominantly been limited to the static (DC) transport and transverse (Hall) channel. 
In this work, we extend the spectral sum rules for frequency-resolved electric conductivity $\sigma (\omega)$ in topological systems, and show that the sum rule for the longitudinal channel is expressed through topological and quantum-geometric invariants. We find that for dispersionless (flat) Chern bands, the  rule is expressed as, 
$ \int_{-\infty}^{+\infty} d\omega \, \Re(\sigma_{xx} + \sigma_{yy}) = C \Delta e^2$, 
where $C$ is the Chern number, $\Delta$ the topological gap, and $e$ the electric charge. In scenarios involving dispersive Chern bands, the rule is defined by the invariant of the quantum metric, and Luttinger invariant, 
$\int_{-\infty}^{+\infty} d\omega \, \Re(\sigma_{xx} + \sigma_{yy}) = 2 \pi e^2 \Delta \sum_{\vec{k}} \Trace \mathcal{G}_{ij}(\vec{k})$+(Luttinger invariant), 
where $\Trace \mathcal{G}_{ij}$ is invariant of 
the Fubini-Study metric (defining spread of Wannier orbitals). We further discuss the physical role of topological and quantum-geometric invariants in spectral sum rules. Our approach is adaptable across varied topologies and system dimensionalities.
\end{abstract}

\maketitle

In quantum matter, topological invariants are fundamental mathematical constructs reflecting the global properties of non-interacting and interacting systems. Tracing back to the TKNN invariant in quantized Hall effect \cite{Thouless1982}, topological and quantum-geometric notions and their experimental observables play a crucial role in advancing quantum matter research, in particular in the directions of unconventional superconductivity \cite{Torma2022}, and topological quantum computation \cite{Nayak2008}. The rising experimental focus on flat-band Chern insulators and their interacting phases, notably in twisted transition metal dichalcogenides
\cite{TMD-FCI-2023a,TMD-FCI-2023b,TMD-FCI-2023c,TMD-FCI-2023d}, accentuates the increasing relevance of topological effects with respect to the dispersive effects, and poses new challenges to the exploration and interpretation of transport data.  
Despite these advances, the frequency-resolved quantum transport and the generalized spectral sum rules remain an under-explored frontier in quantum matter research.

In the seminal paper \cite{Kubo1957}, Kubo pioneered the quantum field theory techniques to compute the conductive properties of solid-state materials.
Among accomplishments of the paper, Kubo formulated the \textit{spectral sum rule} as an invariant of the system, relating the frequency sum (integral) of the real part of AC conductivity $\sigma (\omega)$ with the charge density $n$, 
\begin{align}
\int \limits_{0}^{\infty}   d \omega \  \Re \,  \sigma (\omega)  = \frac{\pi n e^2}{2 m},
\label{sum-rule}
\end{align}
where $e$ and $m$ are electron charge and mass. 
The significance of the sum rule \eqref{sum-rule} is undebatable. Since its inception, it has served as a reliable tool for experimentalists to assess electron density $n$ via frequency-resolved conductivity measurements. 

\begin{figure}[b]
    \centering
    \includegraphics[width=0.7 \columnwidth]{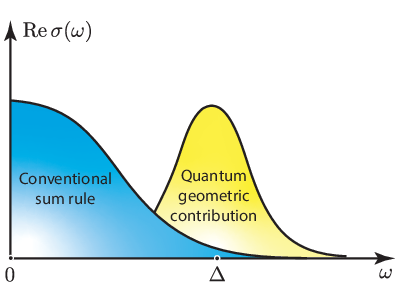}
    \caption{ 
    Schematic illustration of spectral sum rule in topological systems: conventional non-topological intraband contributions (blue) versus quantum-geometric contribution (yellow). }  
    \label{Spectral-illustration}
\end{figure}

The conventional sum rule \eqref{sum-rule}, elegantly derived for noninteracting electrons at absolute zero, retains its form for Fermi gas even upon inclusion of interactions and thermal effects \cite{Kubo1957,Martin1967, Izuyama1973}.
This surprising immunity to interactions becomes evident upon application of Luttinger theorem \cite{Abrikosov1965, Seki2017}, 
\begin{align}
\int \limits_{- \infty}^{\infty}   d \omega \  \Re \,  \sigma (\omega)  = \frac{\pi e^2}{m}   \int _{G^{-1}(0, \vec k) >0} \frac{d^{D} \vec k}{(2 \pi)^{D}}.
\end{align}
where $G(\omega, \vec k)$ is the \textit{interacting} Green's function, and $D$ is dimensionality of the system. The quantity on the right-hand side is the \textit{Luttinger invariant} \cite{Abrikosov1965,Seki2017}. It has a simple interpretation: 
For the interacting Fermi gas, a system with well-resolved  quasiparticle poles,  the Luttinger invariant equates to the volume encompassed by the Fermi surface, an invariant resilient to interaction effects. Henceforth, we will refer to expressions akin to \eqref{sum-rule} as \textit{Luttinger invariant}.

For Bloch-periodic solids, the conventional spectral sum rule \eqref{sum-rule} undergoes transformations, and this may pose several challenges.  
Typically, 
instead of the bare electron mass $m$ in Eq.  \eqref{sum-rule}, one employs the effective mass $\frac{\partial^2 \eps_{n \vec k}}{\partial \vec k^2 }$ of the dispersive band $\eps_{n\vec k}$ \cite{Kubo1957}.
Yet there are known deviations from this procedure, namely in the systems where the effective low-energy description does not presume an effective mass (e.g. for Dirac fermions in monolayer graphene \cite{Gusynin2007, Sabio2008}), or in the systems where interactions modify the nature of quasiparticles (e.g. in superconductors \cite{Hirsch1992,Maiti2010}), which require careful treatment. 
Second, Kubo's original framework explicitly omits the interband effects, inherent to topological systems. 
Third, 
the advent of quantum materials with flat bands presents new challenges and emphasizes the increasing role of quantum geometry of these systems \cite{TKV2019, Kruchkov2023, Torma2022}.  In the case of perfectly flat topological bands, as in chiral models of twisted graphene multilayers \cite{TKV2019,Khalaf2019}, the effective mass becomes infinite, the Fermi surface is undefined, and the Luttinger invariant \eqref{sum-rule} does not seem to be applicable. These factors collectively complicate the application of conventional sum rules in new-generation topological materials, particularly those featuring flat bands.

In this paper, we extend the spectral sum rules to the case of topological systems, with a special focus on those with flat topological bands. We discover that in the systems described by (nearly)-dispersionless electronic bands, the spectral sum rule is defined through the topological and quantum-geometric invariants of the system.
Nontrivial Wannier orbitals in topological systems contribute extra spectral weight due to the interband Berry connections, as depicted in Fig.\ 1. 

Specifically, for the case of two flat Chern bands, we find that the spectral sum rule is given by the \textit{topological invariant} (in this case, the first Chern number $C$),
\begin{align}
\int \limits _{- \infty}^{+\infty}   d \omega \,  \Re (\sigma_{xx} + \sigma_{yy}) = |C| \Delta e^2. 	
\nonumber
\end{align} 
 For the dispersive topological bands, the spectral sum rule is defined by the \textit{quantum metric invariant} (plus the Luttinger invariant),
\begin{align}
  \int \limits _{- \infty}^{+\infty}  d \omega \, \Re (\sigma_{xx} + \sigma_{yy}) & = 2 \pi  e^2 \Delta \sum_{\vec k \in \rm BZ} \Trace \mathcal G_{ij}(\vec k) 
  \nonumber
  \\
 &  + \text{Luttinger invariant}. 
  \nonumber	
\end{align}
These formulations, applicable to weakly interacting topological systems, 
are the key results of our work.
We shall later discuss the physical significance of these invariants. 
Finally, we attribute a topological interpretation to the conventional Kubo sum rule by considering windings of the (interacting) Green’s function.

\section{Methodology}

In this work, we employ quantum transport theory, focusing on fermionic systems with moderate interactions that preserve the order of quasiparticle poles (albeit altering the quasiparticle weight).  
We consider finite temperatures (unless stated otherwise) and use the Matsubara framework for calculations. The central object of our investigation is the current-current correlator (with $\hbar$$=$$1$  henceforth), written in the symmetric form \cite{Kubo1957}
\begin{align}
\Pi_{ij} (\tau) = 	  \langle  \Torder \{ J_{i} (\tau) , J_{j} (0)  \} \rangle	, 
\end{align}
wherein the current operator, expressed in second quantization notations, reads
\begin{align}
\vec J = e \sum_{\vec k \in \rm BZ} c^{\dag}_{\vec k } \boldsymbol{\VV}^{}_{\vec k} 	c^{}_{\vec k} , 
\ \ \ 
\boldsymbol{\VV}^{}_{\vec k} \equiv \frac{\partial \mathcal H}{\partial \vec k} .  
\end{align}
Here $\mathcal H$ is the Hamiltonian of the system, and $e$ is 
the electric charge.  The summation over solid state momenta $\vec k$ encompasses the entire first Brillouin Zone (BZ).
Furtheron, employing Wick's theorem, we arrive to evaluation of quantities 
\begin{align}
\Pi^{\pm}_{ij} (\tau) = e^2 \sum_{\vec k \in \rm BZ}	\Trace \left[ G_{\vec k} (\pm \tau) \VV_i G_{\vec k} ( \tau) \VV_j  \right], 
\label{Pi-tau}
\end{align}
where $\VV_i$ is a shortcut for  $\VV_i  = \partial \mathcal H / \partial k_i$. 
We focus on the two-dimensional (2D) topological materials, and, therefore, spatial indices $i, j$ assume values $x,y$. Although our interest predominantly lies in the longitudinal response (conductivities $\sigma_{xx}$ and $\sigma_{yy}$), we keep different indices $i,j$ for book keeping. 
This distinction will be useful in subsequent discussion on symmetrization procedures.
To evaluate the correlator \eqref{Pi-tau}, we proceed to  Matsubara transform, namely by using expressions 
$
\Pi_{ij} (i \omega_0)  = 
\int _{0}^{\beta}d\tau 
e^{i \omega_0 \tau } 	\Pi_{ij} (\tau)$,   $\Pi_{ij} (\tau)  = \frac{1}{\beta} \sum_{i \omega_0} e^{- i \omega_0 \tau }
	\Pi_{ij} (i \omega_0)
$, 
where $\beta = 1/ T$ is inverse temperature ($k_B$$=$$1$), and $\omega_0$ denotes the external (bosonic) Matsubara frequency.
Upon Matsubara transform, the correlator \eqref{Pi-tau} reads
\begin{align}
	\Pi^{\pm}_{ij} (i \omega_0) = \frac{e^2}{\beta} \sum_{\vec k \in \rm BZ} \sum_{i \omega_n} \Trace [ G_{\vec k} (i \omega_n \pm i \omega_0 ) \VV_i G_{\vec k} ( i \omega_n ) \VV_j ].
\end{align}

We examine the real part of 
the frequency-resolved (complex) conductivity $\sigma (\omega) = \sigma'(\omega) + i \sigma '' (\omega) $. 
Within the framework of Kubo's linear response theory, the real component \( \sigma'(\omega) \) behaves as a symmetric function in frequency. 
Following the analytic continuation of $ \sigma(i \omega_n) = \Pi(i \omega_n)/ i \omega_n $ to the real axis, the real part of the AC conductivity is written through a symmetrized expression, 
\begin{align}
\Re \, \sigma_{ij} (\omega) = \frac{e^2}{4} \sum	_{\vec k \in \rm BZ}  \int \limits_{- \infty}^{+\infty} d \eps  f(\eps)  &  \Trace [ A _{\vec k}  ( {\eps}) V_i  C _{\vec k} ( {\eps; \omega} ) V_j  
\nonumber
\\
+ & (i \leftrightarrow j) ]. 
\label{sigma-symmetric1}
\end{align}
Here $A_{\vec k} (\eps, \vec k) $ is the spectral function and operator $C_{\vec k}$ is given by
\begin{align}
C_{\vec k} ( {\eps; \omega} )  = \frac{(G^R_{\eps - \omega} - G^R_{\eps + \omega} ) + (G^A_{\eps + \omega} - G^A_{\eps - \omega} ) } {i \omega}	
\end{align}
\noindent 
where $G^{\rm A, R}$ represent advanced and retarded  Green's functions. 
The Trace operator to be taken in the band basis. 
It’s important to note that conductivity formula \eqref{sigma-symmetric1} is symmetric under time reversal and properly reproduces the DC limit of longitudinal conductivity ($\omega \to 0 $) \footnote{We remind the reader that we keep only longitudinal indices $i,j$ here. The Hall conductance is given by a similar \textit{antisymmetrized} expression. }.

It is essential to clarify the action of current $\partial \Ham / \partial \vec k $ under the Trace operation.  When the quasiparticles are well-defined, the Bloch bands are well resolved,  and the action of position operator in Bloch basis on the wave function returns both the intraband (diagonal) and interband (off-diagonal) terms \cite{Blount1962}.  
Consequently, in the band basis, we arrive at
 \begin{align}
 \boldsymbol {\mathcal V}_{nm} = \partial_{\vec k}  \eps_{n \vec k} \, \delta_{nm}
 + \Delta_{nm} (\vec k) \, \langle u_{n \vec k}| \partial_{\vec k}  u_{m \vec k}\rangle.
 \label{Bloch velocity}
 \end{align}
 Here $n,m$ represent band indices,  $u_{n \vec k}$ represent Bloch states, and $\Delta_{nm}(\vec k) = \eps_{m \vec k} - \eps_{n \vec k}$ are momentum-resolved band gaps. 
 The first term of \eqref{Bloch velocity} gives the conventional quasiparticle velocity. The second term of \eqref{Bloch velocity}  is an interband effect, proportional to interband Berry connection $\mathcal A_{nm} (\vec k) = - i  \langle u_{n \vec k}| \partial_{\vec k} | u_{m \vec k}\rangle $ and is relevant to topological systems \cite{Haldane2004,Matsuura2010}.

For the subsequent analysis, we define  the multiorbital quantum-geometric tensor $\mathfrak G^{nm}_{ij}$ 
\cite{Ma2010, Matsuura2010,Kruchkov2023}, as given by 
\begin{align}
\mathfrak G^{nm}_{ij}  (\vec k) \equiv \langle \partial_{i} u_{n\vec k} | u_{m\vec k}  \rangle \langle u_{m\vec k} |  \partial_{j} u_{n\vec k}   \rangle .
\label{metric-nm}
\end{align}
Here $\partial_{i}$ denotes momentum derivative $\partial_{k_{i}}$; upper indices $n,m$ correspond to electronic orbitals, while lower indices $i,j$ refer to spatial coordinates $x,y$.
A useful property of multiorbital quantum metric can be obtained upon partial summation over the orbital index,
\begin{align}
\sum_{m\ne n} \mathfrak G^{nm}_{ij}  = \mathfrak G^{(n)}_{ij}	
\label{metric-sum}
\end{align}
where  we have used the resolution of identity, $\sum_{m }  | u_{m \vec k} \rangle \langle u_{m \vec k} | = 1$. 
Here $\mathfrak G^{(n)}_{ij}$ 
is the quantum metric of the $n$-th band, defined by formula
\begin{align}
\mathfrak G^{(n)}_{ij} (\vec k) = \langle \partial_{i} u_{n \vec k} | \left [1-      | u_{n \vec k} \rangle \langle u_{n \vec k}  |  \right] | \partial_{j} u_{n \vec k} \rangle . 
\label{metric}
\end{align}
The quantum metric tensor $\mathfrak G^{(n)}_{ij} (\vec k)$ decomposes into the real part,  
\textit{the Fubini-Study metric} $\mathcal G_{ij} = \Re\, \mathfrak G_{ij}$ \cite{Provost1980}, 
and imaginary part associated with the Berry curvature $\mathcal F_{xy} \epsilon_{ij}= - 2 \Im \, \mathfrak G_{ij}$,
responsible for topology.
Being non-negative and gauge-invariant,  
the quantum metric tensors $\mathfrak G^{(n)}_{ij}$ define 
\textit{quantum distance}.  
As a consequence, there are useful inequalities which relate the real (diagonal) part to the imaginary (off-diagonal) part \cite{Jackson2015}. 

The formalism described here is fundamentally based on the original Kubo formalism \cite{Kubo1957}, but it is refined to a more universal form, enabling to incorporate quantum transport phenomena in \textit{topological systems}.
Specifically, this representation facilitates a comprehensive 
algorithm to address the \textit{interband transitions}, which were neglected in Kubo's original work, but important to systems with nontrivial Wannier orbitals. 
Thereafter, our methodology is suitable for addressing the spectral sum rule for topologically-nontrivial systems.

 We first revisit and rederive the conventional Kubo sum rule \eqref{sum-rule} focusing on topologically-trivial, dispersive bands.  
 We shall see that the sum rule \eqref{sum-rule} is a consequence of the \textit{meromorphicity} of longitudinal conductivity \eqref{sigma-symmetric1}, 
 and 
 we provide a topological interpretation to this statement, insensitive to the strength of moderate interactions. 
 A novel perspective reveals that the Kubo sum rule \eqref{sum-rule} represents the Luttinger invariant, characterized by the winding number of the determinant of interacting Green's functions.  
 After that, we generalize the spectral rule for the topological systems, particularly focusing on the dominant interband contributions in the flat-band limit, and discuss the role of other relevant invariants of the system (in particular, 
 the first Chern number, and Marzari-Vanderbilt invariant $\Trace \mathfrak G$).

\section{Sum rule for topologically-trivial bands}

Firstly, we reexamine the optical sum rule for topologically-trivial bands to rederive Kubo’s 1957 result. However, our approach here differs, based chiefly on the analytical properties of the (interacting) Green's functions without relying on their explicit mathematical form.
In this section, we focus on a single-band electronic system characterized by the dispersion relation 
 $\eps_{\vec k} = \eps (\vec k)$, and
consider additive properties of  $\Trace \sigma_{ij}=(\sigma_{xx} + \sigma_{yy})$.  
Given common crystallographic symmetries, such as $C_3$,  the transport response becomes axis-independent, $\sigma_{xx}=\sigma_{yy}=\sigma$; in this case spatial indices can be dropped. 
For the topologically-trivial band, the only contribution to Bloch velocity \eqref{Bloch velocity} comes from the diagonal term, and the interband effects can be generically neglected. 
Thus Eq.\ \eqref{sigma-symmetric1} simplifies to
\begin{align}
\Re  \, \Trace \sigma_{ij} (\omega) = \frac{e^2}{2}  \sum_{\vec k \in \rm BZ} V^2_{\vec k} \int \limits_{- \infty}^{+\infty} d \eps  f(\eps) \, 
A _{\vec k}  ( {\eps})  C_{\vec k} ( {\eps; \omega} ) . 
\nonumber
\end{align}
Here, $V_{\vec k} \equiv \partial_{\vec k} \eps_{\vec k}$ denotes the conventional quasiparticle velocity in the single Bloch band with dispersion $\eps_{\vec k}$ (first term in Eq.\ \eqref{Bloch velocity}). 
For now, we refrain from making assumptions on the explicit form of propagators $G^{\rm R,A} (\omega)$,  and the specific form of interactions,  assuming that the interactions maintain the quasiparticle poles 
\footnote{We omit the vertex corrections in the electric conductivity analysis. The influence of these corrections can be estimated through the Ward-Takahashi identities \cite{Ward1970,Takahashi1957}, relating the vertex corrections to the self-energy of interacting electrons. In most relevant cases (Ref. \cite{Mahan}, Sec. 8.1.3), this relation can be expressed as $\Gamma_{\vec k, i \omega}  = \left[ 1 - \frac{\partial \Sigma ( i \omega, \vec k) }{\partial (i \omega)} \right] \equiv Z^{-1}_{\vec k, i\omega}$.  For the system with well-resolved quasiparticles, the quasiparticle weight is close to unity, $Z_{\vec k, \eps_{\vec k}} \simeq 1$, hence the vertex corrections present a secondary effect.}.
All information regarding interactions is encapsulated in the spectral function  $A_{\vec k}(\omega)$, which is connected to retarded and advanced propagators as
\begin{align}
G^{R,A}_{\vec k}  (\omega)  =  \int \limits_{-\infty}^{+\infty}   d \omega' \frac{A_{\vec k}(\omega')}{\omega - \omega' \pm i \delta} .
\nonumber
\end{align}
Employing these expressions, and integrating over external frequency $\omega$, one arrives to expression for the frequency sum
\begin{widetext}
\begin{align}
 \int \limits_{-\infty}^{+\infty}   d \omega & \text Re \,  [ \sigma_{xx}(\omega) +  \sigma_{yy}(\omega)]  
 = \frac{e^2}{2 }  \sum_{\vec k \in \rm BZ} V^2_{\vec k} \int \limits_{- \infty}^{+\infty} d \eps  f(\eps) \, 
A _{\vec k}  ( {\eps})  \int \limits_{-\infty}^{+\infty}   d \omega' A_{\vec k}(\omega')  
\nonumber
\\
& \times   \int \limits_{-\infty}^{+\infty}  \frac{d \omega}{i \omega}  \left[
\frac{1}{\eps - \omega - \omega' + i \delta}	
 -
\frac{1}{\eps + \omega - \omega' + i \delta}	  
  +  \frac{1}{\eps + \omega - \omega' - i \delta}	 - \frac{1}{\eps - \omega - \omega' - i \delta}  \right].
\end{align}
\end{widetext}

Subsequent integration over 
$\omega$ can be performed independently of the explicit form of the interacting spectral functions by employing the residue theorem. 
 Indeed, by enclosing the integral contour in the upper half-plane, the pertinent poles are 
$\omega_1 = \omega' - \eps + i \delta$ and $\omega_2 = \eps - \omega' + i \delta$, with residues  $( \omega' - \eps + i \delta)^{-1}$ and $( \omega' - \eps - i \delta)^{-1}$ accordingly.  
Utilizing the Sokhotsky-Plemelj theorem, one arrives to the generic frequency sum rule
 \begin{align}
 \int \limits_{-\infty}^{+\infty}   d \omega  \, \Re [ \sigma_{xx}(\omega) + \sigma_{yy}(\omega) ] 
= -  2 \pi e^2  \sum_{\vec k \in \rm BZ} V^2_{\vec k} \, \,  \mathcal F (\eps_{\vec k} ), 
\label{sum-rule-1}
\end{align}
where we have introduced a momentum-dependent function
 \begin{align}
\mathcal F ( \eps_{\vec k} ) = \int \limits_{- \infty}^{+\infty} d \eps  f(\eps) \, 
  A _{\vec k}  ( {\eps})  \ 
 \fint \limits_{-\infty}^{+\infty}   d \omega' \frac{A_{\vec k} (\omega')}{\eps - \omega'}. 
\end{align}
Notation $ \fint d \omega' ...$ denotes integration in the sense of Cauchy's Principal Value.

Expression \eqref{sum-rule-1} formulates the optical sum rule through subsequent integration with squared velocity, $V^2_{\vec k}= \left( \frac{\partial \eps_{\vec k}}{\partial \vec k} \right)^2$. 
However, this form of summation is not particularly conducive for our objectives. Consequently, we aim to recast the optical sum rule in terms of the \textit{inverse mass tensor},  $m_{ij}^{-1} = \frac{\partial^2 \eps_{\vec k}}{\partial k_i \partial k_j}$, e.g. in the form as 
\begin{align}
\sum_{\vec k \in \rm BZ} \left( \frac{\partial \eps_{\vec k}}{\partial \vec k} \right)^2 \mathcal F (\eps_{\vec k} )   \  \to  \ 
\sum_{\vec k \in \rm BZ } \frac{\partial^2 \eps_{\vec k}}{\partial \vec k^2}  \mathcal W (\eps_{\vec k} )
\end{align} 
To accomplish this transformation, we recognize that the following relationship exists due to the Bloch periodicity,
\begin{align}
\int \limits_{\rm BZ} \frac{d^D {\vec k}}{ (2 
\pi)^D } 
\frac{\partial^2 \eps_{\vec k}}{\partial \vec k^2 }  \, 
\mathcal W (\eps_{\vec k} )    =  - \int \limits_{\rm BZ} \frac{d^D {\vec k}}{ (2 
\pi)^D } 
\left(\frac{\partial \eps_{\vec k} }{\partial \vec k} \right)^2 \frac{ \partial \mathcal W (\eps_{\vec k}) }{\partial \eps_{\vec k}} 
\nonumber
\end{align}  
(the boundary term drops due to the Bloch periodicity of $\eps_{\vec k}$ as an observable). 
Thereafter the unknown function $ W(\eps_{\vec k})$ must fulfill relationship  $\mathcal F (\eps_{\vec k} ) = - { \partial \mathcal W (\eps_{\vec k}) }/{\partial \eps_{\vec k}} $, from which it can be deduced that 
\begin{align}
W(\eps_{\vec k}) = - \int \limits^{\eps_{\vec k}} d E 	\  \mathcal F (E ).
\end{align}
We further examine the isotropic case with $\sigma_{xx} = \sigma_{yy}$, and keep the $xx$ indices. 
In this case, the generic \textit{spectral sum rule} \eqref{sum-rule-1}, conveyed through the tensor of effective mass, now reads as follows:
\begin{align}
 \int \limits_{-\infty}^{+\infty}   d \omega \  \Re [ \sigma_{xx}(\omega) ]   =  2 \pi e^2 \sum_{\vec k \in \rm BZ} \frac{ \mathcal I (\vec k) }{m_{xx} (\vec k)}.
 \label{sum-rule-gen}  
\end{align}
where the numerator is given by expression
\begin{align}
\mathcal I (\vec k) =
 \int d \eps_{\vec k}  
 \int \limits_{- \infty}^{+\infty} d \eps  f(\eps) \,  \fint \limits_{-\infty}^{+\infty}   d \omega'     \frac{  A _{\vec k}  ( {\eps})  A_{\vec k} (\omega')}{\eps - \omega'}	
  \label{int}.
\end{align}
The expression \eqref{sum-rule-gen} maintains a high degree of generality, suitable for (interacting) propagators exhibiting quasiparticle poles. 
Furthermore, it is applicable to \textit{single-band} systems with any form of dispersion relation that are nontopological in nature.
To the best of our knowledge, the spectral sum rule written in terms of propagators in Eq.\ \eqref{sum-rule-gen}, has not been presented in existing literature. 

\textit{Reduction to the Kubo sum rule.}---We shall show in the next steps that the generic spectral sum rule \eqref{sum-rule-gen} converges to the Kubo sum rule \eqref{sum-rule} when applied to nontopological bands with quadratic dispersion.  In the first step, we address a single-band nontopological system characterized by a quasiparticle propagator with the quasiparticle residue $Z_{\vec k}$, and renormalized dispersion $\epsilon_{\vec k}$. This system is characterized by the spectral function
\begin{align}
A_{\vec k} (\omega)	 = \frac{1}{\pi} \frac{\gamma_{\vec k} Z_{\vec k} }{(\omega - \epsilon_{\vec k} + i \gamma_{\vec k}) (\omega - \epsilon_{\vec k} - i \gamma_{\vec k}) }.
\nonumber
\end{align}
In the second step, we evaluate $\omega'$ integral 
in Eq.\ \eqref{int}. 
The residue theorem can be employed for this purpose; the pole at   $\omega' = \eps$ is excluded, and the pole  $\omega' = \epsilon_{\vec k} + i \gamma_{\vec k}$, originating from  $A_{\vec k} (\omega')$, contributes to the evaluation. Consequently, we obtain 
\begin{align}
\mathcal I (\vec k)  = 
 \int d \eps_{\vec k} 
 \int \limits_{- \infty}^{+\infty} d \eps  f(\eps)  A _{\vec k}  ( {\eps}) 
 \frac{ Z_{\vec k}(   
  \eps - \epsilon_{\vec k})}{( \epsilon_{\vec k} - \eps )^2 + \gamma_{\vec k}^2}  .
 \end{align}
 In the third step, the indefinite integral over energies
$\eps_{\vec k}$ is evaluated, 
yielding to an elegant expression 
\begin{align}
\mathcal I (\vec k)   
    = \frac{ Z_{\vec k} n_{\vec k} } {2 } ,
 \end{align}
where $n_{\vec k}$ counts the number of particles encompassed by the Fermi surface, 
 \begin{align}
n_{\vec k} = 
 \int \limits_{- \infty}^{+\infty} d \eps  f(\eps)  A _{\vec k}  ( {\eps}) . 
 \end{align}
In the final step, employing Eq. \eqref{sum-rule-gen}, we derive the optical sum rule for a dispersive, nontopological electronic band, resulting in expression 
   \begin{align}
 \int \limits_{-\infty}^{+\infty}   d \omega  \,  \sigma_{xx}(\omega)  =  \pi e^2 \sum_{\vec k  \in \rm BZ} \frac{Z_{\vec k}  n_{\vec k}}{m_{xx} (\vec k)}.
 \label{sum-rule-dispersive}  
\end{align}
A similar formula can be found in Kubo's original work \cite{Kubo1957}. Note that the \textit{varying} within the Brillouin Zone effective mass 
$m_{xx} (\vec k)$ influences the summation output.

\textit{Interacting Fermi gas}. 
A further refinement of the dispersive Kubo formula \eqref{sum-rule-dispersive} arises in a scenario exhibiting parabolic dispersion and a constant effective mass, represented by $\epsilon_{\vec k} \simeq k^2/ 2 m$ (recalling, $\hbar$$=$$1$), a condition applicable to e.g. interacting Fermi gas. Furthermore, for the uniform Fermi gas the quasiparticle residue is nearly unitary $Z_{\vec k} \simeq 1$. 
In this case,  formula \eqref{sum-rule-dispersive} undergoes simplifications, 
  \begin{align}
  \int \limits_{0}^{+\infty}   d \omega  \,  \Re \sigma_{xx}(\omega)  \simeq   \frac{\pi n e^2 }{2m}. 
  \label{f-sum-conv}
\end{align}
This is precisely the Kubo sum rule 
\eqref{sum-rule} for topologically-trivial systems allowing parabolic approximation.
For a generic band dispersion one should refer to Eq.\ \eqref{sum-rule-dispersive}.

\textit{Topological interpretation of the Kubo sum rule}.  
The immunity of the spectral sum rule against interactions suggests a  topological framing for Eq.\ \eqref{sum-rule}.
Our derivation above, 
different from the original path of Kubo, depends primarily on the \textit{meromorphicity} of the propagators, yet not on their explicit mathematical expressions. The meromorphic properties of Green's functions are generally 
preserved even after inclusion of weak or moderate interactions to the system. 
Consequently, a topological interpretation can be attributed to the Kubo sum rule \eqref{sum-rule}. This interpretation is closely related to \textit{topological interpretation of Luttinger theorem} \cite{Seki2017,Oshikawa2000,Gavensky2023}, which we follow here.  
For simplicity, we consider a single-band system with well-defined Fermi surface, and effective mass $m$.  
In Matsubara formalism, the particle number is given by \cite{Mahan} 
\begin{align}
	n & = \frac{1}{\beta}  \lim_{\tau \to 0^{+}} \sum_{i \omega_n } e^{i \omega_n \tau } \Trace G_{\vec k}(i \omega_n)
	\\
	& = \oint \limits_{\mathcal C}   \frac{dz}{2 \pi i} f(z) \, \, \Trace G_{\vec k}(z)
	\label{particle-number}
\end{align}
where $f(z) = (e^{\beta z}+1)^{-1}$ is the Fermi-Dirac function, and complex contour encompasses singularities of Green's function $G_{\vec k} (z)$, but not the poles of $f(z)$. Further, we would like to evaluate $\Trace G_{\vec k}(z)$ \footnote{Unless stated otherwise, the Trace operation in our notations is always applied to the lower indices of a tensor.}. For this, we use Dyson equation for interacting system 
\begin{align}
	G^{-1}_{\vec k}(z)  = G^{-1}_{0, \vec k} (z)- \Sigma_{\vec k} (z), 
\end{align}
where $G_{0, \vec k} (z) = (z-\mathcal H_0)^{-1}$ is the non-interacting Green's function and $\Sigma_{\vec k} (z)$ is self-energy. 
Taking $z$-derivative on both sides, and multiplying by $G(z)$, one obtains 
\begin{align}
G_{\vec k}(z)  \frac{\partial G^{-1}_{\vec k} (z)}{ \partial z} = G_{\vec k}(z) - G_{\vec k}(z) \frac{\partial \Sigma_{\vec k} (z)}{\partial z}, 
\end{align}
Which upon taking Trace operation leads to 
\begin{align}
\Trace G_{\vec k}(z)  = \Trace  \left[G_{\vec k}(z)  \frac{\partial G^{-1}_{\vec k} (z)}{ \partial z} \right] +   \Trace  \left[ G_{\vec k}(z) \frac{\partial \Sigma_{\vec k} (z)}{\partial z} \right] \nonumber 
\\
 = 
 \frac{\partial}{\partial z}  \ln \det G^{-1}_{\vec k}(z) 
 +   \Trace  \left[ G_{\vec k}(z) \frac{\partial \Sigma_{\vec k} (z)}{\partial z} \right]. \nonumber
\end{align}
Using this expression in \eqref{particle-number}, one obtains
\begin{align}
	n  = &  \oint \limits_{\mathcal C}   \frac{dz}{2 \pi i} f(z) \, \frac{\partial}{\partial z}  \ln \det G^{-1}_{\vec k}(z)
	\\
	+ &
	 \oint \limits_{\mathcal C}   \frac{dz}{2 \pi i} f(z) \,     \Trace  \left[ G_{\vec k}(z) \frac{\partial \Sigma_{\vec k} (z)}{\partial z} \right]. 
	\label{particle-number2}
\end{align}
The Luttinger theorem \cite{Luttinger1960, Abrikosov1965}; see also \cite{Seki2017, Heath2020}, states that when the Fermi surface is well-defined, one has
\begin{align}
\lim_{T \to 0}	n  = \lim_{T \to 0}  \oint \limits_{\mathcal C}   \frac{dz}{2 \pi i} f(z) \, \frac{\partial}{\partial z}  \ln \det G^{-1}_{\vec k}(z) . 
	\label{particle-number3}
\end{align}
Upon altering the integration variables, this result can be rewritten as
\begin{align}
	n  = \frac{1}{2 \pi i}  \oint \limits_{ \det G^{-1}_{\vec k} (z \in  \mathcal C)}   \frac{d \left[ \det G^{-1}_{\vec k} \right] }{\det G^{-1}_{\vec k}}  . 
	\label{particle-number4}
\end{align}
This represents the \textit{winding number} of the determinant of the Green's function around its origin in the $\zeta = \det G^{-1}_{\vec k}$ complex plane \cite{Seki2017}. 
Even after inclusion of moderate interactions,---preserving the structure of the quasiparticles poles,---the winding number remains invariant. 
Consequently, we refer to the Kubo sum rule \eqref{sum-rule} as the \textit{Luttinger invariant}.

\section{Dispersionless topological bands} 
 
 Having revisited the conventional Kubo sum rule \eqref{f-sum-conv},  it's important to highlight the crucial step in this derivation.  It is the \textit{omitting the interband effects}, i.e., the second term in matrix elements,
 \begin{align}
 \boldsymbol {\VV }_{nm} = \partial_{\vec k} \eps_{n \vec k} \, \delta_{nm}
 + \Delta_{nm} (\vec k) \langle u_{n \vec k}| \partial_{\vec k} u_{m \vec k}\rangle. 
 \label{vel-2}
 \end{align}
and retaining the diagonal terms proportional to the quasiparticle velocity in the $n$th band, $\partial_{\vec k} \eps_{n \vec k}$, which leads to Kubo's spectral sum rule \eqref{f-sum-conv}.

In contrast, we now examine a scenario prioritizing the second, interband term in Eq.\ \eqref{vel-2} while omitting the first, dispersive term,
 \begin{align}
 \boldsymbol \VV^{\rm flat}_{nm} =  \Delta_{nm} (\vec k) \langle u_{n \vec k}| \partial_{\vec k} u_{m \vec k}\rangle. 
 \label{velocity-op}
 \end{align}
 This scenario characterizes bands as predominantly flat across the entire Brillouin zone, i.e., $\partial_{\vec k} \eps_{n \vec k} \to 0$. Moreover, the system shows topological characteristics, as the  Berry connection $\mathcal A_{nm}(\vec k)$$\equiv$$-i \langle u_{n \vec k}| \partial_{\vec k} u_{m \vec k}\rangle$ is significant. This set of criteria defines \textit{dispersionless topological states}, which have been recently achieved in moiré heterostructures, including twisted graphene multilayers and twisted transition metal dichalcogenides \cite{TKV2019, Khalaf2019, Wu2019, Wang2022, Ledwith2022a, Kruchkov2022}. In such instances, the multiband Berry connection introduces a nontrivial, momentum-dependent \textit{quantum metric}, which will dominate our discussion in this section.

\subsection{Calculation of the spectral sum rule for perfectly flat bands}

In order to address the effects of quantum metric, it is useful to further symmetrize expression for the real part of frequency-resolved conductivity \eqref{sigma-symmetric1}.  Indeed, using invariance of the Trace under circular shifts, one can rewrite the same expression in the fully-symmetrized form, 
\begin{align}
\Re \, \sigma_{ij} (\omega) = &  \frac{e^2}{8} \sum	_{\vec k \in \rm BZ}  \int \limits_{- \infty}^{+\infty} d \eps  f(\eps)  \Trace \left[ A _{\vec k}  ( {\eps}) V_i  C _{\vec k} ( {\eps; \omega} ) V_j 
\right. 
 \nonumber
 \\
 + & \left.
   C _{\vec k} ( {\eps; \omega} ) \VV_i   A _{\vec k}  ( {\eps}) \VV_j + (i \leftrightarrow j)\right] . 
\label{sigma-symmetric2}
\end{align}
This symmetrization facilitates subsequent index operations.  

Using the symmetrized form \eqref{sigma-symmetric2}, and formula \eqref{velocity-op}, one can take trace in the band basis, leading to 
\begin{widetext} \begin{align}
 \Re \, \sigma_{ij} (\omega) = - \frac{e^2}{8} \sum_{\vec k \in \rm BZ} \sum_{n,m}    \Delta_{nm}^2 (\vec k) \int \limits_{- \infty}^{+\infty} d \eps  f(\eps) 
 \left\{ \left[ A_{n}  ( \eps ,\vec k)  C_{m}  ( {\eps; \omega}; \vec k ) + A_{m}  ( \eps ,\vec k)  C_{n}  ( {\eps; \omega}; \vec k )  \right]   \right. 
 \nonumber 
 \\
  \times   \left.  \langle u_{n \vec k}| \partial_{i} | u_{m \vec k}\rangle
   \langle u_{m \vec k}| \partial_{j} | u_{n \vec k}\rangle
  + (i \leftrightarrow j) \right\}  . 
  \label{sigma-flat-1}
\end{align}
Here  $A_n(\varepsilon;\vec{k})$, $C_{m}  ( {\eps; \omega}; \vec k $) denote diagonal elements of matrices
$A_{{\vec k}}(\varepsilon)$, $C_{\vec k}(\varepsilon; \omega)$
in the band basis. 
The entity in the second line of Eq.\ \eqref{sigma-flat-1} is the multiband quantum geometric tensor $\mathfrak G^{nm}_{ij} (\vec k)$, hence
 \begin{align}
 \Re \, \sigma_{ij} (\omega) = - \frac{e^2}{8} \sum_{\vec k \in \rm BZ} \sum_{n,m}    \Delta_{nm}^2 (\vec k) \int \limits_{- \infty}^{+\infty} d \eps  f(\eps)  
 \left\{ \left[ A_{n}  ( \eps ,\vec k)  C_{m}  ( {\eps; \omega}; \vec k ) + A_{m}  ( \eps ,\vec k)  C_{n}  ( {\eps; \omega}; \vec k )  \right]   
 \left[ \mathfrak G^{nm}_{ij} (\vec k) + \mathfrak G^{nm}_{ji} (\vec k) \right]
  \right\}  . 
  \label{sigma-flat-2}
\end{align}
This represents the most refined expression of AC conductivity via the quantum metric.
\end{widetext}

\noindent
Employing Eq.\ \eqref{sigma-flat-2}, we can calculate the spectral sum for dispersionless topological bands. Specifically, we find 
\begin{align}
\Re \int \limits_{-\infty}^{+\infty}   d \omega   \,  \Trace   \sigma_{ij}(\omega) 
 = -  \frac{ e^2}{4}  \sum_{\vec k}  \sum_{nm} F_{nm}  \Delta_{nm}^2 
(\vec k)   
  \Trace \mathcal G^{nm}_{ij} (\vec k),
\label{sum-flat}
\end{align} 
where Trace operation on both sides of equation acts only on lower indices ($i,j$),
and tensor $ F_{nm}$ is given by
\begin{align}
 F_{nm}  =   \int \limits_{- \infty}^{+\infty} d \omega  \int \limits_{- \infty}^{+\infty} d \eps  f(\eps) 
& \left[   A_{n}  ( \eps ,\vec k)  C_{m}  ( {\eps; \omega}; \vec k )  \right. 
  \nonumber 
 \\
 +  & \left. A_{m}  ( \eps ,\vec k)  C_{n}  ( {\eps; \omega}; \vec k ) \right]. 
\end{align} 
Using  expression \begin{align}
C_{m} ( {\eps; \omega}; \vec k )  & = \frac{G^R_m ({\eps - \omega}, \vec k)  - G^R_m  ({\eps + \omega}; \vec k ) } {i \omega}	
\nonumber 
\\
& +  \frac{G^A_m ({\eps + \omega}, \vec k) - G^A_m ({\eps - \omega} , \vec k) } {i \omega}	
\end{align}
we arrive to expression 
 \begin{align}
 F_{nm}  =  & \int \limits_{- \infty}^{+\infty} d \omega'  \int \limits_{- \infty}^{+\infty} d \eps  f(\eps) 
 A_{n}  ( \eps ,\vec k)  A_{m}  ( \omega'; \vec k )  I (\omega', \eps) 
 \nonumber
 \\
 + &  (n \leftrightarrow m). 
\end{align} 
where 
\begin{align}
I (\omega', \varepsilon) &  =   \int \limits_{- \infty}^{+\infty}   \frac{d \omega}{i \omega} 
 \left[
\frac{1}{\eps - \omega - \omega' + i \delta}	
 -
\frac{1}{\eps + \omega - \omega' + i \delta}	 
\right. 
\nonumber
\\
&  +  \left. \frac{1}{\eps + \omega - \omega' - i \delta}	 - \frac{1}{\eps - \omega - \omega' - i \delta}  \right]
  \nonumber
  \\ 
&  =  
 \frac{2 \pi}{ \omega' - \eps + i \delta} +  \frac{2 \pi}{ \omega' - \eps - i \delta} . 
 \label{I-int}
\end{align} 
In the final line of Eq.\ \eqref{I-int}, we have utilized the residue theorem. Subsequently, by invoking the Sokhotsky-Plemelj theorem, one finds that tensor $F_{nm}$, defining the sum rule, is simplified to  
 \begin{align}
 F_{nm}  =   4 \pi   \int \limits_{- \infty}^{+\infty} d \eps  f(\eps) 
 A_{n}  ( \eps ,\vec k)   \fint \limits_{-\infty}^{+\infty}     \frac{A_{m}  ( \omega'; \vec k )  d \omega' }{ \eps - \omega' }  +  (n \leftrightarrow m) .
 \label{f-flat}
\end{align} 

The combination of formulas \eqref{sum-flat} and symmetric $\vec k$-dependent tensor \eqref{f-flat} represents the most general expression for the spectral sum rule in dispersionless topological states,  and serves as a cornerstone result of our work. 
Below we show  how this formula simplifies in  the "clean limit."

\textit{Clean limit}.---Considering a noninteracting system at an arbitrary temperature, the spectral function simplifies to a delta function,  $A^{n}_{\vec k}  ( {\eps}) =  \delta (\eps - \eps_{n \vec k})$.
This, in turn, leads to explicit expression for the tensor $F_{nm}$,
\begin{align}
F_{nm}  &   =  
 4 \pi  f(\eps_{n \vec k}) \, 
   \fint \limits_{-\infty}^{+\infty}   d \omega'   \frac{ \delta (\omega' - \eps_{m \vec k}) }{\eps_{n \vec k} - \omega'} + (m \leftrightarrow n)
    \nonumber
  \\
 &    = 4 \pi   \frac{f(\eps_{n \vec k}) } {\eps_{n \vec k} - \eps_{m \vec k} } + (m \leftrightarrow n) 
 =  \frac{4 \pi [f (\eps_{n \vec k})  -  f (\eps_{m \vec k})] }{\Delta_{mn} (\vec k)} .
 \nonumber
\end{align}
Using this expression in Eq.\ \eqref{sum-flat}, we obtain the sum for the clean limit of perfectly flat bands as
 \begin{align}
\Re  \int \limits_{-\infty}^{+\infty}  & d \omega  \, \Trace [ \sigma_{ij}(\omega) ]
  =    \pi e^2   \sum_{\vec k  \in \rm BZ}   \sum_{nm}  f_{nm}  \Delta_{nm}  (\vec k)   \, 
 \Trace  \mathcal G^{nm}_{ij} (\vec k).
\label{sum-flat-2}
\end{align}  
 where $f_{nm} = f (\eps_{n \vec k})  -  f (\eps_{m \vec k})$ is the difference in Fermi-Dirac distribution functions. 
 Equation \eqref{sum-flat-2} applies to a noninteracting system and incorporates temperature dependence through function $f_{nm} (T) $. Generally, this contribution \eqref{sum-flat-2} is non-vanishing. A straightforward demonstration follows with a selected example below.

\textit{Two Flat Chern Bands at Zero Temperature}.--- Consider a prototypical topological system consisting of two flat electronic bands distinguished by their  Chern invariants $\pm C$, and separated by a gap of value $\Delta$. These bands fulfill the  "trace condition" \cite{Kruchkov2022, Roy2014},  
\begin{align}
	\Trace \mathcal G_{ij}(\vec k) = |\mathcal F_{xy} (\vec k)| ,
 \label{trace condition}
\end{align}
relating their quantum metric trace with the magnitude of the Berry curvature $F_{xy}$.  Assuming the lower band is fully occupied, the upper band is vacant, and  $\mathcal F_{xy} >0$ for the lowest band,  we derive the spectral sum rule
 \begin{align}
 \int \limits_{-\infty}^{+\infty}   d \omega  \, \Re [ \sigma_{xx}(\omega) + \sigma_{yy}(\omega) ] 
& =  2  \pi e^2 \Delta  \sum_{\vec k  \in \rm BZ}  
\Trace \mathcal G_{ij} (\vec k) 
\nonumber
\\
& =  2  \pi e^2 \Delta  \sum_{\vec k  \in \rm BZ}  
\mathcal F_{xy} (\vec k),
\nonumber
\end{align}
or simply
\begin{align}
 \int \limits_{-\infty}^{+\infty}   d \omega  \, \Re [ \sigma_{xx}(\omega) + \sigma_{yy}(\omega) ] 
 = C e^2 \Delta. 
\end{align}  
In the last line, we have used the definition of the (first) Chern number
   \begin{align}
 C = \frac{1}{2 \pi} \int_{\rm BZ}    d  \vec k \ \mathcal F_{xy}. 
 \end{align}
Therefore, we find that in the perfectly flat bands, the spectral sum rule is expressed through the topological invariants of the system, in this case the first Chern number. 
For a symmetric system where $\sigma_{xx}= \sigma_{yy} =\sigma$, we have
\begin{align}
 \int \limits_{0}^{+\infty}   d \omega  \, \Re \sigma (\omega)  
 = \frac{C e^2 \Delta}{4}. 
\end{align}

 \section{Spectral sum rules for dispersive topological systems}

In the prior sections, we have explored two distinct scenarios and their underlying assumptions leading to the spectral sum rules.  Starting from the equation 
 \begin{align}
 \boldsymbol {\VV }_{nm} = \partial_{\vec k} \eps_{n \vec k} \, \delta_{nm}
 + \Delta_{nm} (\vec k) \langle u_{n \vec k}| \partial_{\vec k} | u_{m \vec k}\rangle, 
 \label{vel-3}
 \end{align}
 which defines the lattice currents, we have derived: 
(a)  For dispersive nontopological bands (Sec.\ II), neglecting the second term in \eqref{vel-3} resulted in the conventional spectral sum rule defined by the Luttinger invariant.
(b)  For dispersionless topological bands (Sec.\ III), omitting the first term in \eqref{vel-3} has led us to the spectral sum rule characterized by the topological invariants.
 We now aim to unify these two disparate cases by examining the spectral sum rule for \textit{dispersive topological bands}.  
 
\textit{Gauge invariance.}--- At first glance, integrating cross-terms from \eqref{vel-3} might yield terms linear in the interband Berry connection, $\mathcal A_{nm}$, a quantity which is not gauge-invariant. On the other hand, the frequency-resolved conductivity $\sigma ( \omega)$, being a physical observable, \textit{must} maintain gauge invariance. 
Previous sections sidestepped this discussion by either (i) omitting $\mathcal A_{nm}$ in Sec.\ II, or (ii) including it in gauge-invariant combinations in Sec.\ III. 
To clarify, let us rewrite the expression AC conductivity \eqref{sigma-symmetric3} using the properties of Trace operation, 
\begin{align}
\Re \, \sigma_{ij} (\omega) =  &  \frac{e^2}{4}  \sum	_{\vec k \in \rm BZ}  \sum_{m,n } \int \limits_{- \infty}^{+\infty} d \eps  f(\eps) 
\nonumber   
\\
 & \times [ A_{n}   \mathcal V^{(i)}_{nm}  C_{m} \mathcal V^{(j)}_{mn}
+   A_{m}   \mathcal V^{(i)}_{mn}  C_{n} \mathcal V^{(j)}_{nm}] . 
\label{sigma-symmetric3}
\end{align}
A direct substitution of 
Eq.\ \eqref{vel-3} in Eq.\ \eqref{sigma-symmetric3} makes evident that gauge-dependent cross-terms cancel each other. 
Consequently, neither the frequency-resolved conductivity nor the spectral sum rule depends on cross-terms containing linear in $\mathcal A_{nm}$ terms. The outcome remains \textit{gauge-invariant}, as anticipated.  
Owing to gauge invariance, 
formula Eq.\ \eqref{sigma-symmetric3}, has two separate contribions, $\mathcal O [(\partial_{\vec k} \eps_{n \vec k})^2]$  and $ \mathcal O [ \mathcal A_{nm}^2]$.  We refer to these additive terms as \textit{intraband} and \textit{interband} contributions.

\textit{Intraband contributions.---} We first start from the intraband contributions. In agreement with our previous derivation \eqref{sum-rule-1}, we obtain
    \begin{align}
 \int \limits_{-\infty}^{+\infty}   d \omega  \,  \Trace \sigma_{ij}(\omega)  =  \pi e^2 \sum_{\vec k  \in \rm BZ}  \sum_{m} 
Z_{\vec k} n_{\vec k}   \, \Trace  [\partial_{i} \partial_{j} \eps_{m \vec k}], 
\end{align}
 where the trace is taken in $i,j$ basis.  
When the Fermi level resides within the gap, the formula at $T=0$ in the clean limit simplifies to 
\begin{align}
\Re  \int \limits_{-\infty}^{+\infty}   d \omega  \,  \Trace \sigma^{\rm intra}_{ij}(\omega)  =  \pi e^2 \sum_{\vec k  \in \rm BZ}  \left. \sum_{m} \right. ^{'} 
 \Trace  [\partial_{i} \partial_{j} \eps_{m \vec k}], 
 \label{intra}
\end{align}
with the prime over the sum indicating summation over occupied bands. 
This contribution precisely aligns with the Kubo formula of 
Ref.\ \cite{Kubo1957}, which omits the interband processes. 
In the insulating phase, the intraband term \eqref{intra}  vanishes.

 \textit{Interband contributions.---}  We now incorporate the interband processes from 
Eq.\ \eqref{vel-3} to frequency-resolved conductivity \eqref{sigma-symmetric3}.  This procedure is analogous to derivations described in Sec.\ III,  which gives  
 \begin{align}
\Re   \int \limits_{-\infty}^{+\infty}   d \omega  
\, \Trace  \sigma^{\rm inter }_{ij}(\omega)  
=   \pi e^2 &  \sum_{\vec k  \in \rm BZ}   \sum_{nm}  [f_{nm}  \Delta_{nm}  (\vec k)  ] 
\nonumber
\\
&  \times  \Trace \mathcal G^{nm}_{ij} (\vec k).
\end{align}

Consequently, the \textit{generalized spectral sum rule} for topological systems combines both the conventional (dispersive) and quantum-geometric contributions, represented by  
\begin{align}
\Re   \int \limits_{-\infty}^{+\infty}   d \omega  
\, \Trace  \sigma_{ij}(\omega)  
=  \pi e^2 \sum_{\vec k  \in \rm BZ}  \sum_{m} 
\left\{  n_{\vec k}   \, \Trace  [\partial_{i} \partial_{j} \eps_{m \vec k}] \right . 
\nonumber 
\\
+   \sum_{n \ne m}  
\left.  [f_{nm}  \Delta_{nm}  (\vec k)  ] 
 \Trace \mathcal G^{nm}_{ij} (\vec k) \right\}, 
 \label{sum-rule-gen-CI}
\end{align} 
 where we assume quasiparticles well-defined, i.e., with $Z_{\vec k} \approx 1$. 
The generalized sum rule \eqref{sum-rule-gen} for dispersive topological systems, while not explicitly referencing topological invariants, relates to quantum-geometric and Luttinger invariants. 
The link of Eq.\ \eqref{sum-rule-gen} to the topological invariants, such as Chern numbers, manifests in the spectral inequalities \textit{(Appendix A)}.

\textit{Two flat Chern bands at zero temperature.}--- 
Examining a two-band Chern insulator with the Fermi level placed within the topological gap, the dispersive term in Eq.\ \eqref{sum-rule-gen-CI} drops out, and the important term, incorporating the interband quantum metric effects, significantly simplifies. 
Due to the inherent symmetry of the two-band Chern insulator, we have $\mathcal G^{(1)}_{ij} (\vec k)  = \mathcal G^{(2)}_{ij} (\vec k) \equiv \mathcal G_{ij} (\vec k)$, and $f_{12} \Delta_{12} = f_{21} \Delta_{21}$. Furtheron, by employing partial summation identity \eqref{metric-sum}, we arrive to the sum rule  
\begin{align}
 \Re   \int \limits_{-\infty}^{+\infty}   d \omega  
\, \Trace  \sigma_{ij}(\omega)  
=  2 \pi e^2   
 \sum_{\vec k  \in \rm BZ}   \Delta  (\vec k)  \,  
 \Trace \mathcal G_{ij} (\vec k)  , 
 \label{sum-rule-CI}
\end{align} 
where 
$\Trace \mathcal G_{ij} (\vec k)$ is the Fubini-Study metric of either band. 
This sum rule is reminiscent of the recent finding of integrated current noise in topological systems, which is proportional to $  
 \sum_{\vec k }   \Delta^2  (\vec k)  \,  
 \Trace \mathcal G_{ij} (\vec k) $ \cite{KRYU2023}. 
 This attributes to a similar internal structure of the integrands in Kubo quantum transport theory.

Consideration for the relevant crystallographic symmetries, such as $C_3$, leads to symmetric conductivity tensors $\sigma_{xx}(\omega)$$=$$\sigma_{yy}(\omega)$$= $$\, \sigma(\omega)$. Therefore, the sum rule for the Chern insulator \eqref{sum-rule-CI} further simplifies as
\begin{align}
    \int \limits_{0}^{+\infty}   d \omega  \, 
\Re \sigma (\omega)  
=   \frac{\pi e^2 }{2} 
 \sum_{\vec k  \in \rm BZ}   \Delta  (\vec k)  \,  
 \Trace \mathcal G_{ij} (\vec k)  .  
 \label{sum-rule-CI-2}
\end{align} 
This formulation echoes the original Kubo sum rule \eqref{sum-rule} upon assigning 
$m_{*}= \left[n \sum_{\vec k}   \Delta  (\vec k)  \,  
 \Trace \mathcal G_{ij} (\vec k) \right]^{-1}$, where $m_* \ne 0$  is a parameter possessing the dimensionality of mass.

\textit{Quantum-geometric invariant.}   
Quantum-geometric entity $\Trace \mathcal G_{ij} (\vec k)$, 
entering sum rules for 
$\int _{0}^{+\infty}   d \omega  \, 
\Re[  \sigma (\omega) ]$ 
and explicitly in 
$\int _{0}^{+\infty}   d \omega  \, 
\Re [\sigma (\omega)]/ \omega$ 
(and, more generically, in sum rules  of form $
    \int _{0}^{+\infty}   d \omega  \,  \omega^n \, 
\Re [ \sigma (\omega)] $), 
has a specific meaning as the gauge-invariant part of Wannier spread functional \cite{Marzari1997}. Its physical interpretation relates to the overlap of electronic orbitals in topological systems, connecting to the expectation values of the squared position operator.
Indeed, by introducing the Wannier functions $\mathcal W^{(n)}_{\vec R}$ by virtue of relation. $| u_{n \vec k} \rangle \equiv \sum_{\vec R} e^{i \vec k \cdot \vec R} | \mathcal W^{(n)}_{\vec R} \rangle $, one can show that the spread of electronic orbitals
\begin{align}
\Omega = \sum_{n} \left[ \langle r^2_n \rangle - \langle r_n \rangle^2 \right],
\end{align}
can be decomposed into two terms
   \begin{align}
\Omega = \Omega_I + \tilde \Omega. 
\end{align}
Both terms are positive-definite; however, only  term 
$\Omega_I$ is gauge-invariant. Marzari and Vanderbilt demonstrated that this gauge-invariant component can be represented as  \cite{Marzari1997}, 
\begin{align}
 \Omega_I  = \sum_{\vec k \in \rm BZ} \Trace \mathcal G_{ij} (\vec k), 
 \label{MV-inv}
\end{align}
which corresponds to the Brillouin-zone-averaged quantum metric. Moreover, $\Trace \mathcal G (\vec k)$ is by definition one of the invariants of quantum metric. We henceforth refer to invariant \eqref{MV-inv} as the \textit{Marzari-Vanderbilt invariant}. 
This invariant underpins the formulation of maximally-localized Wannier orbitals \cite{Marzari2012}.
Notably, in topological systems, invariant $\Omega_I$ cannot take zero values due to topological constraints that hinder the construction of maximally-localized Wannier orbitals \cite{Thouless1984,Brouder2007}.    Consequently, the lower bound of the invariant 
$\Omega_I$ is bounded from below by topological invariants, ensuring substantial orbital overlap \cite{Kruchkov2022}.

The sum rules for topological systems of type \eqref{sum-rule-CI-2} can be physically interpreted through the role of Marzari-Vanderbilt invariant.  
As quantum-geometric invariant $\Trace \mathcal G (\vec k)$ is responsible for finite and significant overlap in electronic orbitals in topological bands, electrons in such system become strongly entangled. The resulting AC currents are strongly influenced by the degree of electronic overlap $\Omega_I$, which result into the sum rules \eqref{sum-rule-CI-2}:  
The larger is the orbital overlap, the larger is the spectral sum rule invariant in topological systems. Finally, in dispersionless Chern bands,  $\Omega_I$ up to a dimensional prefactor coincides with the Chern number $C$ \cite{Kruchkov2022}, thus offering simple physical interpretation to the sum rule invariant $\equiv C e^2 \Delta$   in Sec.\ III.

\section{Overview}

In this study, we have refined the spectral sum rules for topological systems. 
We have found that unique topological features, paired with nontrivial quantum geometry, introduce previously overlooked terms in the spectral sum rules. 
Traditional derivations, starting from Kubo's pioneering work \cite{Kubo1957}, largely focused on dispersive, topologically-trivial systems, neglecting the interband contributions to electric conductivity.
Yet, for topological fermions, the interband contributions culminate into the  sum rule altered by the term
  \begin{align}
  \pi e^2  \sum_{\vec k  \in \rm BZ}   \sum_{m,n}  
  f_{nm}  \Delta_{nm}  (\vec k)  \,  
 \Trace \mathcal G^{nm}_{ij} (\vec k) ,   
\nonumber
\end{align}
with $\mathcal G^{nm}_{ij}$ representing the underlying quantum metric. Given the nascent exploration of quantum metric in quantum materials,  it is understandable that such the quantum-geometric effects in sum rules have been overlooked for years.  Existing conductivity formulas had not been addressing quantum-geometric effects.

\begin{figure}[b]
    \centering
    \includegraphics[width = 0.7 \columnwidth]{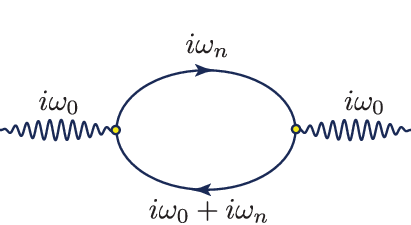}
    \caption{Current-current correlator. Solid lines represent the interacting propagator, wiggly lines denote insertion of the current operator,  vertex corrections are not shown (footnote [27]). The diagrams naturally embed invariants of the system. For the interacting Fermi liquid, this invariant is Luttinger invariant, represented by the conserved $U(1)$ charge.}  
    \label{diagram}
\end{figure}

Moreover, quantum-geometric effects on spectral sum rules often simplify to fundamental invariants: the Luttinger invariant (Sec.\ II), topological invariants (Chern number in Sec.\ III), and the quantum-geometric invariants (Marzari-Vanderbilt invariant in Sec.\ IV). Notably, even the conventional Kubo sum rule \eqref{sum-rule} permits a topological reinterpretation, explaining its well-established robustness against interactions from a fresh perspective.

Our detailed analysis underscores that spectral sum rules inherently convey system invariants. In diagrammatic terms (see Fig.\ \ref{diagram}), transport diagrams upon external frequency summation embed information on the invariants of the fermionic system. For interacting Fermi liquids, these diagrams reveal Luttinger invariant, i.e., the conservation of $U(1)$ charge, as was established by Kubo and subsequent research. In dispersive topological systems, the diagrams with topological currents reveal additional \textit{Marzari-Vanderbilt} invariant, responsible for the overlap  of current-carrying electronic orbitals. In a dispersionless topological state, the transport diagrams contain the only possible invariant, the system-specific topological charge.

Historically, topological invariants have been predominantly probed using quantum Hall measurements, shot noise, and local probes \cite{Hasan2010}.  Notably, our findings suggest that the spectral sums measurement can serve as a different tool for direct probing topological and quantum-geometric invariants,   and provide further information on the topological band gaps \cite{KRYU2023}, without usage of Hall-bar experimental platforms.

\section*{Appendix A: Geometric inequalities, band gaps, and topological invariants}

\textbf{Mathematical inequalities}. 
Since quantum geometric tensor is nonnegatively-definite, there is a number of useful mathematical inequalities between its real part (Funini-Study metric $\mathcal G$) and imaginary part (Berry curvature $\mathcal F$). 
In particular, the "determinant inequality" reads \cite{Jackson2015, Roy2014} 
 \begin{align}
 4 \det \mathcal G_{ij} (\vec k)  \geq |\mathcal F_{xy} ( \vec k) |^2 .
 \end{align}
This inequality sets the lower bound on \textit{quantum volume}.

The second, stricter inequality addresses relation between the trace of quantum metric and Berry curvature modulus,
\begin{align}
\Trace \mathcal G_{ij} (\vec k)  \geq |\mathcal F_{xy} ( \vec k) |. 
\end{align}
This inequality is saturated in ideal flat Chern bands, leading to the so-called "trace condition" \eqref{trace condition}; 
see further discussions in Refs.\ \cite{Kruchkov2022, Ledwith2023}. As a consequence, the gauge-invariant part of Marzari-Vanderbilt invariant \eqref{MV-inv} of ideal flat Chern bands is quantized \cite{Kruchkov2022}. 
In dimensionless units, one obtains
\begin{align}
 \Omega^{\text{flat}}_I  = \sum_{\vec k \in \rm BZ} \Trace \mathcal G_{ij} (\vec k) =  |C| / 2 \pi. 
 \label{MV-inv-2}
\end{align}

\textbf{Souza, Wilkens and Martin (SWM) formula.}
Previously, Souza, Wilkens and Martin \cite{Souza2000} have established the following sum rule 
 \begin{align}
 \int \limits_{0}^{+\infty}   d \omega  \, \frac{\Re [ \sigma_{xx}(\omega) + \sigma_{yy}(\omega) ] }{ \omega}
 = \pi e^2 \sum_{\vec k \in \rm BZ }  
\Trace \mathcal G_{ij} (\vec k) . 
\label{Souza}
 \end{align}
Note that there is an extra factor of $1/\omega$ in the integrand. As a consequence, expression \eqref{Souza} does not contain the band gap $\Delta$.
This expression agrees with the results of Sec.\ IV and the sum rule for current noise \cite{KRYU2023}. Comparing these results, the $n$-th moment of the spectral sum rule can be qualitatively summarized as
 \begin{align}
\Re  \int \limits_{0}^{+\infty}   d \omega  \, \omega^n \, \sigma(\omega)
\propto \Delta^{n+1} \sum_{\vec k \in \rm BZ}  
\Trace \mathcal G_{ij} (\vec k) 
\nonumber
\\+ \text{[dispersive terms, if any]}. 
 \end{align}
In particular, the spectral sum rule for the current noise returns second power of the topological band gap, $|\Delta|^2$, as found in Ref.\ \cite{KRYU2023}.

Furtheron, let us discuss a special inequality proposed by Souza, Wilkens and Martin, \cite{Souza2000}. Consider for simplicity a noninteracting two-orbital Chern insulator characterized by band gap $\Delta (\vec k)$. For simplicity, consider also symmetry $\sigma_{xx} = \sigma_{yy} = \sigma$. Then formula \eqref{Souza} can be bounded from above 
\begin{align}
\int \limits_{0}^{+\infty}   d \omega  \, \frac{\Re   \sigma(\omega)  }{ \omega}
\leq  \frac{ \int \limits_{0}^{+\infty}   d \omega  \, \Re \sigma (\omega)  } {\Delta_0} . 
\end{align}
This inequality is based on the observation that \textit{typically} (yet not necessarily always 
\footnote{For example, in case of strong interactions violating the quasiparticle picture, this is not obvious.}), 
the \textit{ideal insulators} do not show optical absorption below $\omega = \Delta_0 $, where $\Delta_0 = \text{min} \Delta (\vec k)$. Hence, for the noninteracting case one obtains that
\begin{align}
 \int \limits_{0}^{+\infty}   d \omega  \, \frac{\Re  \sigma(\omega)  }{ \omega}
\leq  & \frac{\pi e^2 }{2 \Delta_0} 
 \sum_{\vec k  \in \rm BZ}   \Delta  (\vec k)  \,  
 \Trace \mathcal G_{ij} (\vec k)  
 \\
 +  & \frac{\pi e^2}{2 \Delta_0} \sum_{\vec k  \in \rm BZ}  \sum_{m}  f_{m \vec k}  
\partial^2_{\vec k} \eps_{m \vec k}   .
\label{estimate}
\end{align} 
Replacing now the left-hand side with SWM equality \eqref{Souza}, one obtains 
\begin{align}
\frac{|C|}{2 \pi} \leq \sum_{\vec k}  
\Trace \mathcal G_{ij} (\vec k) 
\leq  & \frac{1 }{ \Delta_0} 
 \sum_{\vec k  \in \rm BZ}   \Delta  (\vec k)  \,  
 \Trace \mathcal G_{ij} (\vec k)  
 \\
 +  & \frac{1}{\Delta_0} \sum_{\vec k  \in \rm BZ}   \sum_{m}  f_{m \vec k} 
\partial^2_{\vec k} \eps_{m \vec k} 
\label{estimate2}
 \end{align} 
or for the bound on the gap
\begin{align}
\Delta_0  & \leq  \frac{2 \pi}{|C|}  \sum_{\vec k  \in \rm BZ}   \Delta  (\vec k)  \,  
 \Trace \mathcal G_{ij} (\vec k) 
\label{ineq1}
 \\
 & + \frac{2 \pi}{ |C|}  \sum_{\vec k  \in \rm BZ}   \sum_{m}  f_{m \vec k}   
\partial^2_{\vec k} \eps_{m \vec k} . 
\label{ineq2}
 \end{align} 
This inequality involves bounds on band gaps. 
A similar expression, yet without quantum-geometric term \eqref{ineq1}, has been recently introduced in Ref.\ \cite{Onishi2023}.  
The saturation of inequality \eqref{ineq1}+\eqref{ineq2}
takes place for perfectly flat topological bands. 
For such systems, the band gap is unbounded in single-particle formalism.

In conclusion, incorporating the quantum metric in the spectral sum rules is important. Oversights in this aspect can influence conclusions regarding the quantum transport properties of the system. Flat bands are very sensitive to this contribution.

\bibliography{Refs}

\end{document}